\documentclass[12pt]{article} \usepackage[dvips]{color,graphicx} \usepackage{amsmath,amssymb}
\DeclareGraphicsExtensions{.eps}

\newcommand{\ap}{\ensuremath{\alpha'}} % Inverse string tension
\newcommand{\ls}{\ensuremath{l_s}} %String length
  \def\p{\partial} 
  \def\jnodot{{\mbox{\emph{\j}}}}
 \newcommand{\tr}{\mathop{\rm Tr}} 
\def\expec#1{\langle #1 \rangle}  
 \def\slash#1{\ensuremath{\;/\!\!\!\! #1}}

\newcommand{\cF}{{\mathcal{F}}}  
\newcommand{\cL}{\mathcal{L}}  \newcommand{\cN}{{\mathcal{N}}}
\newcommand{\cO}{{\mathcal{O}}}

\newcommand{\bS}{{\mathbf{S}}}  

\newcommand{\tret}{{t_{\mbox{\scriptsize ret}}}}

\newcommand{\tx}{{\tilde{x}}}
\newcommand{\ttau}{{\tilde{\tau}}}
\newcommand{\tPi}{{\tilde{\Pi}}}
\def\sring#1{{\;\mathring{}\!\!#1}} \def\dring#1{{\:\mathring{}\;\mathring{}\!\!\!#1}}
\def\tring#1{{\mathring{}\:\mathring{}\:\mathring{}\!\!\!#1}}
\def\qring#1{{\mathring{}\,\mathring{}\,\mathring{}\,\mathring{}\!\!\!\!\,#1}} %Document format
\title{\bf A Tail of a Quark in $\cN=4$ SYM} \author{Mariano Chernicoff\footnote{e-mail:
mariano@nucleares.unam.mx}, J.~Antonio Garc\'{\i}a\footnote{e-mail: garcia@nucleares.unam.mx} ~and
Alberto G\"uijosa\footnote{e-mail: alberto@nucleares.unam.mx} \\{\small Departamento de F\'{\i}sica
de Altas Energ\'{\i}as, Instituto de Ciencias Nucleares}\\ {\small Universidad Nacional Aut\'onoma de
M\'exico}\\ {\small Apdo. Postal 70-543, M\'exico D.F. 04510}} \date{}

\begin{document}
\maketitle
\begin{abstract} We study the dynamics of a `composite' or `dressed'
quark in strongly-coupled large-$N_c$ $\cN=4$ super-Yang-Mills, making use of the AdS/CFT
correspondence. We show that the standard string dynamics nicely captures the physics of the quark
and its surrounding non-Abelian field configuration, making it possible to derive a relativistic
equation of motion that incorporates the effects of radiation damping. {}From this equation one can
deduce a non-standard dispersion relation for the composite quark, as well as a Lorentz covariant
formula for its rate of radiation. We explore the consequences of the equation in a few simple
examples. \end{abstract}

%\vspace{3cm}

%\tableofcontents

\section{Introduction and Summary}

When a charge radiates, energy conservation dictates that it must be subjected to a reactive force
originating from its self-field, that tends to damp its motion. In the context of classical
electrodynamics, the study of this damping or radiation reaction force began over a century ago
\cite{abraham,lorentz,schott,dirac}, and continues to this day \cite{ghw,harte}. Reviews and
additional references on the subject may be found in \cite{rohrlich,jackson,yaghjian,poisson}.

In a non-relativistic approximation, the dynamics of an electron that is modeled as a vanishingly
small spherically symmetric charge distribution is controlled by the Abraham-Lorentz equation
\cite{abraham,lorentz} \begin{equation}\label{al}
m\left(\ddot{\vec{x}}-t_e\dddot{\vec{x}}\right)=\vec{F}~, \end{equation} where $\dot{}\equiv d/dt$
and $t_e\equiv 2e^2/3mc^3$ is a timescale set by the classical electron radius. In this equation, the
damping force (the second term in the left-hand side) is seen to be proportional to the jerk
$\vec{\jnodot}\equiv \dot{\vec{a}}\equiv \ddot{\vec{v}}$. The search for a Lorentz-covariant version
of (\ref{al}) led to the (Abraham-)Lorentz-Dirac equation \cite{dirac}, \begin{equation}\label{ald}
m\left(\dring{x^{\mu}}-t_e\left[ \tring{x^{\mu}}-{1\over c^2}\dring{x_{\nu}}\,
\dring{x^{\nu}}\,\sring{x^{\mu}}\right]\right)=\cF^{\mu}~, \end{equation} with $\;\mathring{}\equiv
d/d\tau$, $\tau$ the proper time (defined such that
$\;\mathring{}\!\!x^{\mu}\;\mathring{}\!\!x_{\mu}=-c^2$) and
$\cF^{\mu}\equiv\gamma(\vec{F}\cdot\vec{v}/c,\vec{F})$ the four-force. The second term within the
square brackets (proportional to the square of the proper acceleration) is the negative of the rate
at which energy and momentum is carried away from the charge by radiation, according to the covariant
Lienard(-Larmor-Heaviside-Abraham) formula. So, strictly speaking, it is only this term that can
properly be called radiation reaction. The first term within the square brackets, usually called the
Schott term, and whose spatial part yields the damping force of (\ref{al}) in the non-relativistic
limit, is known to arise from the effect of the charge's `near' or `bound' (as opposed to radiation)
field \cite{teitelboim,rohrlich}.

The appearance of third-order terms in (\ref{al}) and (\ref{ald}) leads to unphysical behavior,
including pre-accelerating and self-accelerating (or `runaway')  solutions. These deficiencies are
known to originate from the assumption that the charge is pointlike.\footnote{This assumption leads
to the further complication of an infinite electromagnetic self-energy, which has already been been
absorbed within the renormalized mass $m$ shown in (\ref{al}) and (\ref{ald}). An alternative
approach which circumvents this divergence has been proposed recently in \cite{ghw}.} For a charge
distribution of small but finite size $l$, the above equations can be shown to be truncations of
expressions that involve an infinite number of derivatives $(l d/dt)^n$ (and generally include
 terms that are non-linear in these derivatives), but are physically sound as long as $l>ct_e$ and
 \cite{rohrlich,jackson,yaghjian}.

Of course, one should keep in mind that the unphysical behavior implied by (\ref{al}) and (\ref{ald})
would be visible only for time and distance scales smaller than the Compton wavelength
$\lambda_C\equiv\hbar/m$ of the charge, and thus lies outside of the actual range of validity of
classical electrodynamics. How the preceding story generalizes to the case of fully quantum
electrodynamics (QED) has been studied from different angles in
\cite{monizsharp,johnson,martin,rosenfelder} and references therein. In particular, in
\cite{monizsharp} it was shown that, for a pointlike non-relativistic electron, QED leads to an
equation of motion with an infinite number of higher derivatives, implying that the electron acquires
an effective size $l=\lambda_C$ due to its surrounding cloud of virtual particles.

Going further to non-Abelian gauge theories is a serious challenge.\footnote{See respectively
\cite{classicalYMrad} and \cite{weakYMrad} for work on radiation within classical and
(weakly-coupled) quantum Yang-Mills theory.} %missing refs? Nevertheless, it is the purpose of this
paper to show that the AdS/CFT correspondence \cite{malda,gkpw,magoo} allows us to examine this
question rather easily in quantum strongly-coupled non-Abelian gauge theories. The essence of the
matter is that in the context of this duality the quark corresponds to the tip of a string, whose
body codifies the profile of the non-Abelian (near and radiation) fields sourced by the quark. In
other words, the quark has a tail, and it is this tail that is responsible for the damping force.
Indeed, this mechanism has already been seen at work in the recent computations of the drag force
exerted on the quark by a thermal plasma, which is described in dual language in terms of a string
living on a black hole geometry \cite{hkkky,gubser}. Our analysis makes it clear that, irrespective
of whether a spacetime black hole is present or not, the body of the string plays the role of an
energy sink, as befits its identification as the embodiment of the gluonic degrees of
freedom.\footnote{On the other hand, energy loss via the string does turn out to be closely
associated with the appearance of a \emph{worldsheet} horizon,
  as noticed initially in \cite{gubserqhat,ctqhat} at finite temperature and emphasized in
  \cite{dragtime} for the zero temperature case. This association has been further studied in
  \cite{dominguez,xiao,beuf,nolineonthehorizon}.}

We expect this basic story to apply generally to all examples of the gauge/string duality, including
cases with finite temperature or chemical potentials, but for simplicity we will concentrate on the
case of quark motion in the vacuum of $\cN=4$ super-Yang-Mills (SYM), where, building upon previous
work \cite{mikhailov,dragtime}, we can achieve full analytic control. As has been remarked several
times in the past, it is interesting that even in this non-confining theory the gluonic field
configuration can be encoded in a `QCD' string, albeit one that lives in a curved higher dimensional
spacetime. Our results are a direct consequence of this amazing fact.

The paper is organized as follows. In Section \ref{strings2quarkssec}, we explain how the standard
string dynamics is mapped by the AdS/CFT correspondence onto the dynamics of an extended radiating
particle. We begin by setting up our problem in Section \ref{basicsubsec}, reviewing the relevant
context and emphasizing some key features---most notably, the fact that the quark with finite mass
that the correspondence puts at our disposal is automatically `dressed' or `composite', as discussed
around Eq.~(\ref{Fsquared}). We then attack the problem in Section \ref{eomsubsec}, where we derive
an equation of motion for this quark, Eq.~(\ref{eom}), which constitutes our main result.\footnote{A
brief report of this derivation was given in the recent letter \cite{lorentzdirac}.} As explained at
length in the paragraphs that follow it, this equation is a nonlinear generalization of (\ref{ald}),
which incorporates the effects of radiation damping, but has no pre-accelerating or self-ac
 celerating solutions. From this equation one can read off a non-standard dispersion relation for the
 quark, Eq.~(\ref{pq}), as well as a Lorentz covariant formula for its radiation rate,
 Eq.~(\ref{radiationrate}). An interesting novel feature in these expressions is their dependence on
 the external force exerted on the quark, which is a reflection of its extended, and hence
 deformable, nature. We close Section \ref{strings2quarkssec} by commenting on our failure to rewrite
 (\ref{eom}) in terms of an action principle.

In Section \ref{examplessec}, we explore some of the physics implied by (\ref{eom}), specializing to
a few simple examples. For the case of one-dimensional motion, we show in Section \ref{onedimsubsec}
that even though, as expected on physical grounds, zero external force implies zero acceleration, the
converse is not true: the quark will not accelerate when subjected to an external force that takes
the specific form (\ref{constantvelocity}) (which is identically zero only when the parameter
$t_0\to\pm\infty$). More generally, for each given quark trajectory there is a one-parameter family
of possible external forces. This again is a manifestation of the fact that, because of the extended
character of the quark, the energy supplied to it can not only increase its velocity, but also modify
its associated gluonic field profile. We end the paper by studying the nonrelativistic limit of
(\ref{eom}) in Section \ref{nrsubsec}, where the linearized form of the expressions allows us to make
direct contact with the energy analysis of (\cite{mikhailov,dragtime}) and to easily write down an
action principle.

All in all, then, we have in (\ref{eom}) a physically sensible and interesting description of the
dynamics of a composite quark in $\cN=4$ SYM. This result serves, on the one hand, to illustrate the
power of the AdS/CFT correspondence, and on the other, to shed some light on the largely uncharted
terrain of radiation in strongly-coupled non-Abelian gauge theories. It would be interesting to
extend this analysis in a number of directions. In particular, it seems worthwhile to explore the
manner in which the split between intrinsic and radiated energy (and momentum) of the quark achieved
in \cite{mikhailov,dragtime} and the present paper, via examination of the string worldsheet,
manifests itself in the gluonic field profile, by directly computing the expectation value of the
energy-momentum tensor or similar local operators \cite{dampingfield}. It is also natural to try to
carry over some of the present methods to the finite temperature context \cite{dampingtemp}, where
one sh
 ould be able to make contact with previous AdS/CFT analyses of energy loss in a thermal plasma (a
 rather large body of work detonated by the seminal works \cite{hkkky,gubser,ct,liu}), including the
 interesting recent studies of Brownian motion \cite{rangamani,sonteaney,iancu}.

\section{From Strings to Quarks}\label{strings2quarkssec} \subsection{Basic setup}
\label{basicsubsec} It is by now well-known that strongly-coupled $\cN=4$ $SU(N_c)$ SYM with coupling
$g_{YM}$ is dual to Type IIB string theory on a background that asymptotically approaches the
AdS$_5\times\bS^5$ geometry\footnote{From this point on we work in natural units $c=1=\hbar$.}
\begin{eqnarray}\label{metric} ds^2&=&G_{MN}dx^M dx^N={R^2\over z^2}\left(
-dt^2+d\vec{x}^2+{dz^2}\right)+R^2 d\Omega_5~, \\ {R^4\over \ls^4}&=&g_{YM}^2
N_c\equiv\lambda~\nonumber \end{eqnarray} (with a constant dilaton and $N_c$ units of Ramond-Ramond
five-form flux through the five-sphere), where $\ls$ denotes the string length \cite{malda}. The
radial direction $z$ is mapped holographically into a variable length scale in the gauge theory
\cite{uvir}. The directions $x^{\mu}\equiv(t,\vec{x})$ are parallel to the AdS boundary $z=0$ and are
directly identified with the gauge theory directions. The state of IIB string theory described by the
unperturbed metric (\ref{metric}) corresponds to the vacuum of the $\cN=4$ SYM theory, and the closed
string sector describing (small or large) fluctuations on top of it fully captures the gluonic ($+$
\emph{adjoint} scalar and fermionic) physics.

{}From the gauge theory perspective, the introduction of an open string sector associated with a
stack of $N_f$ D7-branes
 in the geometry
(\ref{metric}) is equivalent to the addition of $N_f$ hypermultiplets in the \emph{fundamental}
representation of the $SU(N_c)$ gauge group, breaking the supersymmetry down to $\cN=2$ \cite{kk}.
These are the degrees of freedom that we refer to as `quarks,' even though they include both spin
$1/2$ and spin $0$ fields.  For $N_f\ll N_c$, the backreaction of the D7-branes on the geometry can
be sensibly neglected; in the field theory this corresponds to working in a `quenched' approximation
which disregards quark loops (as well as the positive beta function they would generate). The
D7-branes cover the four gauge theory directions, and extend along the radial AdS direction up from
the boundary at $z=0$ to a position $z=z_m$ where they `end' (meaning that the $\bS^3\subset\bS^5$
that they are wrapped on shrinks down to zero size), which is inversely proportional to the quark
mass, \begin{equation}\label{zm} z_m={\sqrt{\lambda}\over 2\pi m}~. \end{equation}

An isolated quark is dual to an open string that extends radially from the D7-branes to the AdS
horizon at $z\to\infty$. The string dynamics follows as usual from the Nambu-Goto action
\begin{equation}\label{nambugoto} S_{\mbox{\scriptsize NG}}=-{1\over 2\pi\ap}\int
d^2\sigma\,\sqrt{-\det{g_{ab}}}\equiv \int d^2\sigma\,\cL_{\mbox{\scriptsize NG}}~, \end{equation}
where $g_{ab}\equiv\p_a X^M\p_b X^N G_{MN}(X)$ ($a,b=0,1$) denotes the induced metric on the
worldsheet. In our work the entire string will be taken (consistently with the corresponding
equations of motion) to lie at the `North Pole' of the $\bS^5$ (the point where the
$\bS^3\subset\bS^5$ that the D7-branes are wrapped on collapses to zero size), so the angular
components of the metric (which are associated with the orientation of the gauge theory fields in the
internal $SU(4)$ symmetry group) will not play any role, and the lower endpoint of the string will
necessarily lie at $z=z_m$.

We can exert an external force $\vec{F}$ on the string endpoint by turning on an electric field
$F_{0i}=F_i$ on the D7-branes. This amounts to adding to the Nambu-Goto action the usual minimal
coupling \begin{equation} \label{couplingtoA} S_{\mbox{\scriptsize F}}=\int
d\tau\,A_{\mu}(X(\tau,z_m))\p_{\tau}X^{\mu}(\tau,z_m)~, \nonumber \end{equation} or, in terms of the
quark worldline, \begin{equation}\label{externalforce} S_{\mbox{\scriptsize F}}=\int
d\tau\,A_{\mu}(x(\tau))\,\sring{x^{\mu}}(\tau)~. %=\int d\tau\,\cF_{\mu}(\tau)x^{\mu}(\tau)
\end{equation}

Notice that the string is being described (as is customary) in first-quantized language, and, as long
as it is sufficiently heavy, we are allowed to treat it semiclassically. In gauge theory language,
then, we are coupling a first-quantized quark to the gluonic ($+$ other SYM) field(s), and then
carrying out the full path integral over the strongly-coupled field(s) (the result of which is
codified by the AdS spacetime), but treating the path integral over the quark trajectory
$x^{\mu}(\tau)$ in a saddle-point approximation.

Variation of the string action $S_{\mbox{\scriptsize NG}}+S_{\mbox{\scriptsize F}}$ implies the
standard Nambu-Goto equation of motion for all interior points of the string, plus the standard
boundary condition \cite{acny} \begin{equation}\label{stringbc}
\Pi^{z}_{\mu}(\tau)|_{z=z_m}=\cF_{\mu}(\tau)\quad\forall~\tau~, \end{equation} where
\begin{equation}\label{pizmu} \Pi^{z}_{\mu}\equiv \frac{\p\cL_{\mbox{\scriptsize NG}}}{\p(\p_z
X^{\mu})} ={\sqrt{\lambda}\over 2\pi}\left(\frac{(\p_{\tau}X)^2 \p_z X_{\mu}-(\p_{\tau}X\cdot \p_z
X)\p_{\tau}X_{\mu}}{z^2\sqrt{(\p_{\tau}X\cdot \p_z X)^2-(\p_{\tau}X)^2(1+(\p_z X)^2)}}\right)
\end{equation} is the worldsheet (Noether) current associated with spacetime momentum, and we haver
recognized $\cF_{\mu}=-F_{\nu\mu}\p_{\tau}x^{\nu} =(-\gamma\vec{F}\cdot\vec{v},\gamma\vec{F})$ as the
Lorentz four-force.

For the interpretation of our results it will be crucial to keep in mind that the quark described by
this string is not `bare' but `composite' or `dressed'. This can be seen most clearly by working out
the expectation value of the gluonic field surrounding a static quark located at the
origin\footnote{More precisely, the operator in the left-hand side of (\ref{Fsquared}) is the dual of
the dilaton field, and includes not only the standard Yang-Mills term but also scalar and fermion
contributions that can be found in \cite{ktr}, and which we suppress for notational simplicity.}
\cite{martinfsq}, \begin{equation}\label{Fsquared} {1\over 4 g_{YM}^2}\expec{\tr F^2(x)}
={\sqrt{\lambda}\over 16\pi^2|\vec{x}|^4}\left[1-\frac{1+{5\over 2}\left({2\pi
m|\vec{x}|\over\sqrt{\lambda}}\right)^2}{\left(1+\left({2\pi
m|\vec{x}|\over\sqrt{\lambda}}\right)^2\right)^{5/2}}\right]~. \end{equation} For $m\to\infty$
($z_m\to 0$), this is just the Coulombic field expected (by conformal invariance) for a pointlike
charge. For finite $m$ the profile is still Coulombic far away from the origin but in fact becomes
non-singular at the location of the quark, \begin{equation} {1\over 4 g_{YM}^2}\expec{\tr
F^2(x)}={\sqrt{\lambda}\over 128\pi^2}\left[15\left({2\pi m\over\sqrt{\lambda}}\right)^4-{35\over
|\vec{x}|^4}\left({2\pi m|\vec{x}|\over\sqrt{\lambda}}\right)^6+\ldots\right]~
\;\mbox{for}~\;|\vec{x}|< {\sqrt{\lambda}\over 2\pi m}~.\nonumber \end{equation}
 As seen in these equations, the characteristic thickness of this non-Abelian charge distribution
 is precisely the length scale $z_m$ defined in (\ref{zm}). This is then the size of
the gluonic cloud that surrounds the quark, or in other words, the analog of the Compton wavelength
for our non-Abelian source.

It is interesting to note that the string can be alternatively viewed as a Born-Infeld string, i.e.,
a soliton of the gauge and scalar fields on the D7-brane \cite{callanmalda}. Since the small
fluctuations of these fields (corresponding to microscopic open strings) are known to be dual to
mesons, the composite quark itself can be thought of as a soliton constructed by aligning a large
number of mesons \cite{mateos}. The cloud surrounding our quark is  then best thought of as `mesonic'
rather than `gluonic'. Mesons are indeed
 known to be
 the lightest states in the spectrum of the strongly-coupled gauge theory, with masses of order
 $m_{\mbox{\scriptsize mes}}\equiv 1/z_m= 2\pi m/\sqrt{\lambda}\ll m$ \cite{martinmeson},
and form factors with size set by $z_m$ \cite{strassler}.

So, to summarize, $z_m$ can properly be called the \emph{quark} Compton wavelength insofar as it
gives the size of the cloud of virtual particles surrounding the quark, but one should bear in mind
that it is given not by $1/m$ but by $1/m_{\mbox{\scriptsize mes}}$, and in this sense it could also
be referred to as the \emph{meson} Compton wavelength.

\subsection{Equation of motion for the quark} \label{eomsubsec}

The first analysis of an accelerating quark via the AdS/CFT correspondence was carried out in
\cite{cg}, which used tools developed in \cite{dkk} to study the dilatonic waves given off by small
fluctuations on a radial string in AdS$_5$, and infer from them the profile of the gluonic field
$\expec{\tr F^2(x)}$ in the presence of a quark undergoing small oscillations. The results of
\cite{cg} painted an interesting picture of the propagation of nonlinear waves in $\cN=4$ SYM, but
did not allow a definite identification of waves with the $1/|\vec{x}|$ falloff associated with
radiation. (More recently, this falloff has been successfully detected in the same setup through a
calculation of the energy-momentum tensor $\expec{T_{\mu\nu}}$ \cite{mo}.)

In \cite{cg} it was noted that for $z_m\to 0$ and in the linearized approximation, the string action
(\ref{nambugoto}) correctly implies the expected action for an ordinary non-relativistic particle of
mass $m\to\infty$. In the present section we will obtain the relativistic generalization of this
result retaining the full non-linear structure of the Nambu-Goto string, and then further extend the
analysis to the case of finite $m$.

We will take as our starting point the results obtained in a remarkable paper by Mikhailov
\cite{mikhailov}, which we now briefly review (a more detailed explanation can be found in
\cite{dragtime}). This author considered an infinitely massive quark, and was able to solve the
equation of motion for the dual string on AdS$_5$, for an \emph{arbitrary} timelike trajectory of the
string endpoint. In terms of the coordinates used in (\ref{metric}), his solution is
\begin{equation}\label{mikhsol} X^{\mu}(\tau,z)=z{dx^{\mu}(\tau)\over d\tau}+x^{\mu}(\tau)~,
\end{equation} with $x^{\mu}(\tau)$ the worldline of the string endpoint at the AdS boundary--- or,
equivalently, the worldline of the dual, infinitely massive, quark--- parametrized by its proper time
$\tau$.

Combining (\ref{metric}) and (\ref{mikhsol}), the induced metric on the worldsheet is found to be
\begin{equation}\label{wsmetric} g_{\tau\tau}={R^2\over
z^2}(z^2\,\mathring{}\;\mathring{}\!\!\!x^2-1),\qquad g_{zz}=0,\qquad g_{z\tau}=-{R^2\over z^2},
\end{equation} implying in particular that the constant-$\tau$ lines are null, a fact that plays an
important role in Mikhailov's construction. The solution (\ref{mikhsol}) is `retarded', in the sense
that the behavior at time $t=X^{0}(\tau,z)$ of the string segment located at radial position $z$ is
completely determined by the behavior of the string endpoint at an \emph{earlier} time $\tret(t,z)$
obtained by projecting back toward the boundary along the null line at fixed $\tau$. An analogous
`advanced' solution built upon the same endpoint/quark trajectory can be obtained by reversing the
sign of the first term in the right-hand side of (\ref{mikhsol}). In gauge theory language, this
choice of sign corresponds to the choice between a purely outgoing or purely ingoing boundary
condition for the waves in the gluonic field at spatial infinity. Both on the string and the gauge
theory sides, more general configurations should of course exist, but obtaining them explicitly is
difficult due to the highly non-linear character of the system. Henceforth we will focus solely on
the retarded solutions, which are the ones that capture the physics of present interest, with
influences propagating outward from the quark to infinity.

{}From the $\mu=0$ component of (\ref{mikhsol}), parametrizing the quark worldline by $x^0(\tau)$
instead of $\tau$, and using $d\tau=\sqrt{1-\vec{v}^{\,2}}dx^0$, where $\vec{v}\equiv d\vec{x}/dx^0$,
the relation that defines the retarded time follows as \begin{equation}\label{tret}
t=z{1\over\sqrt{1-\vec{v}^{\,2}}}+\tret~, \end{equation} where the endpoint velocity $\vec{v}$ is
meant to be evaluated at $\tret$. In these same terms, the spatial components of (\ref{mikhsol}) can
be formulated as \begin{equation}\label{xmikh}
\vec{X}(t,z)=z{\vec{v}\over\sqrt{1-\vec{v}^{\,2}}}+\vec{x}(\tret)=(t-\tret)\vec{v}+\vec{x}(\tret)~.
\end{equation} Using (\ref{tret}) and (\ref{xmikh}), Mikhailov was able to rewrite the total string
energy in the form \begin{equation}\label{emikh} E(t)={\sqrt{\lambda}\over
2\pi}\int^t_{-\infty}d\tret
\frac{\vec{a}^{\,2}-\left[\vec{v}\times\vec{a}\right]^2}{\left(1-\vec{v}^{\,2}\right)^3}
+E_q(\vec{v}(t))~, \end{equation} where of course $\vec{a}\equiv d\vec{v}/dx^0$. The first term
codifies the accumulated energy \emph{lost} by the quark over all times prior to $t$, and is
surprisingly seen to have precisely the same form as the standard Lienard formula from classical
electrodynamics.\footnote{Possible experimental implications of this result have been explored in
\cite{kharzeev}.} The second term in the above equation arises from a total derivative on the string
worldsheet, and gives the expected Lorentz-covariant expression for the energy intrinsic to the quark
\cite{dragtime}, \begin{equation}\label{edr} E_q(\vec{v})={\sqrt{\lambda}\over
2\pi}\left.\left({1\over\sqrt{1-\vec{v}^{\,2}}}{1\over z}\right)\right|^{z_m=0}_{\infty}=\gamma m~.
\end{equation} For the spatial momentum, \cite{mikhailov,dragtime} similarly find
\begin{equation}\label{pmikh} \vec{P}(t)={\sqrt{\lambda}\over 2\pi}\int^t_{-\infty}d\tret
\frac{\vec{a}^{\,2}-\left[\vec{v}\times\vec{a}\right]^2}{\left(1-\vec{v}^{\,2}\right)^3}
\vec{v}+\vec{p}_q(\vec{v}(t))~, \end{equation} with \begin{equation}\label{pdr}
\vec{p}_q={\sqrt{\lambda}\over 2\pi}\left.\left({\vec{v}\over\sqrt{1-\vec{v}^{\,2}}}{1\over
z}\right)\right|^{z_m=0}_{\infty}=\gamma m\vec{v}~. \end{equation} We see then that, in spite of the
non-linear nature of the system, Mikhailov's procedure leads to a clean separation between the tip
and the tail of the string, i.e., between the quark (including its near field) and its gluonic
radiation field. We will now exploit this separation to study in more detail the dynamics of the
quark.

Our initial observation is that, when we regard the Nambu-Goto action as a functional of the quark
trajectory $x^{\mu}$ by plugging (\ref{wsmetric}) back into (\ref{nambugoto})$+$(\ref{couplingtoA}),
we can explicitly carry out the integral over $z$ to obtain
\begin{eqnarray}\label{relativisticparticle} S_{\mbox{\scriptsize NG}}+S_{\mbox{\scriptsize
F}}&=&-{R^2\over 2\pi\ap}\int d\tau \int_{z_m\to 0}^{\infty} {dz\over z^2}+\int
d\tau\,A_{\mu}(x(\tau))\,\sring{x^{\mu}}(\tau)\\ {}&=& -m\int d\tau~+\int
d\tau\,A_{\mu}(x(\tau))\,\sring{x^{\mu}}(\tau)~,\nonumber \end{eqnarray} which is evidently the
standard action for a pointlike externally forced relativistic particle (with mass $m\to\infty$).
Notice that the associated equation of motion does \emph{not} include a damping force, which is just
as one would expect for an infinitely massive charge, because the coefficient $t_e\propto 1/m$ of the
damping terms in (\ref{al}) and (\ref{ald}) approaches zero as $m\to\infty$.

Let us now consider the more interesting case of a quark with finite mass, $z_m>0$, where, as we
emphasized in the previous subsection, our non-Abelian source is no longer pointlike but has size
$z_m$. In this case the string endpoint is at $z=z_m$, and we must again require it to follow the given quark trajectory, $x^{\mu}(\tau)$. As before, this condition by itself does not pick out a unique string embedding. Just like we discussed for the infinitely massive case below (\ref{wsmetric}), we additionally require the solution to be `retarded' or `purely outgoing', in order to focus on the gluonic field causally set up by the quark. As in \cite{dragtime}, we can inherit this structure by truncating a suitably selected retarded Mikhailov solution.
The embeddings of
interest to us can thus be regarded as the $z\ge z_m$ portions of the solutions (\ref{mikhsol}),
which are parametrized by data at the AdS boundary $z=0$.\footnote{That this direct truncation indeed retains the retarded structure of the solutions is manifestly confirmed in our final rewriting (\ref{mikhsolzm}), where information is seen to propagate upward along the body of the string, i.e., from the UV to the IR of the gauge theory.} Henceforth we will use tildes to label
these (now merely auxiliary) data, and distinguish them from the actual physical quantities
(velocity, proper time, etc.) associated with the endpoint/quark at $z=z_m$, which will be denoted
without tildes.

In this notation, (\ref{mikhsol}) reads \begin{equation}\label{mikhsoltilde}
X^{\mu}(\ttau,z)=z{d\tx^{\mu}(\ttau)\over d\ttau}+\tx^{\mu}(\ttau)~. \end{equation} Repeated
differentiation of this equation with respect to $\ttau$ and evaluation at $z=z_m$ (where we can read
off the quark trajectory $x^{\mu}(\ttau)\equiv X^{\mu}(\ttau,z_m)$) leads to the recursive relations
\begin{eqnarray}\label{recursion} {d x^{\mu}\over d\ttau}&=&z_m{d^2\tx^{\mu}\over
d\ttau^2}+{d\tx^{\mu}\over d\ttau}~,\nonumber\\ {d^2 x^{\mu}\over d\ttau^2}&=&z_m{d^3\tx^{\mu}\over
d\ttau^3}+{d^2\tx^{\mu}\over d\ttau^2}~,\nonumber\\ {}&\vdots&{}\nonumber\\ {d^n x^{\mu}\over
d\ttau^n}&=&z_m{d^{n+1}\tx^{\mu}\over d\ttau^{n+1}}+{d^n\tx^{\mu}\over d\ttau^n}~. \end{eqnarray}
Adding these equations respectively multiplied by $(-z_m)^{n-1}$, we can deduce that
\begin{equation}\label{recursionsolution1} {d \tx^{\mu}\over d\ttau}={d x^{\mu}\over d\ttau}-z_m{d^2
x^{\mu}\over d\ttau^2}+z^2_m{d^3 x^{\mu}\over d\ttau^3}-\ldots~, \end{equation} and, upon further
differentiation, \begin{equation}\label{recursionsolution2} {d^2 \tx^{\mu}\over d\ttau^2}={d^2
x^{\mu}\over d\ttau^2}-z_m{d^3 x^{\mu}\over d\ttau^3}+z^2_m{d^4 x^{\mu}\over d\ttau^4}-\ldots~.
\end{equation} This last expression already takes us halfway towards the equation we are after, but
we still need to find a relation between $d\ttau$ and the endpoint/quark proper time $d\tau$, and
similarly rewrite $d^2 \tx^{\mu}/d\ttau^2$ in terms of quantities at the actual string boundary
$z=z_m$ instead of the auxiliary data at $z=0$.

The first task is easy: from (\ref{mikhsoltilde}) it follows that $$ dX^{\mu}=dz{d \tx^{\mu}\over
d\ttau}+d\ttau\left(z{d^2\tx^{\mu}\over d\ttau^2}+{d \tx^{\mu}\over d\ttau}\right)~, $$ which
evaluated at fixed $z=z_m$ implies $$ dx^{\mu}=d\ttau\left(z_m{d^2\tx^{\mu}\over d\ttau^2}+{d
\tx^{\mu}\over d\ttau}\right)~, $$ and therefore \begin{equation}\label{tau} d\tau^2\equiv
-dx^{\mu}dx_{\mu}=d\ttau^2\left[1-z^2_m\left({d^2\tx\over d\ttau^2}\right)^2\right]~. \end{equation}
To arrive at this last equation, we have made use of the fact that $\ttau$ is by definition the
proper time for the auxiliary worldline at $z=0$, so $(d \tx/d\ttau)^2=-1$ and $(d
\tx/d\ttau)\cdot(d^2 \tx/d\ttau^2)=0$.

What remains then is to express $d^2 \tx^{\mu}/d\ttau^2$ as a function of quark data. For this we
note first that, upon substituting the solution (\ref{mikhsoltilde}), the momentum current
(\ref{pizmu}) (with appropriate tildes) evaluated at $z=z_m$ simplifies to
\begin{equation}\label{pitilde} {2\pi\over\sqrt{\lambda}}\tPi^{z}_{\mu}={1\over z_m}{d^2
\tx_{\mu}\over d\ttau^2}+\left({d^2\tx\over d\ttau^2}\right)^2{d\tx_{\mu}\over d\ttau}~.
\end{equation} To avoid possible confusion, we should note that the tilde in the left-hand side does
not indicate evaluation at $z=0$ (as all other tildes do), but the fact that this current is defined
as charge (momentum) flow per unit $\ttau$. The corresponding flow per unit $\tau$ is clearly
just\footnote{The transformation rules for $\Pi^{a}_{\mu}$ under more general reparametrizations can
be found in, e.g., \cite{dragqqbar}.} $\Pi^{z}_{\mu}=(\p\ttau/\p\tau)\tPi^{z}_{\mu}$, and it is this
object which according to (\ref{stringbc}) must equal the external force $\cF_{\mu}$. Using this,
(\ref{tau}) and the first equation of (\ref{recursion}) in (\ref{pitilde}), one can deduce (after
some straightforward algebra) that \begin{equation}\label{atilde} {d^2 \tx_{\mu}\over
d\ttau^2}=\frac{1}{\sqrt{1-z^4_m \slash{\cF}^2}}\left(z_m \slash{\cF}_{\mu}-z^3_m \slash{\cF}^2{d
x_{\mu}\over d\tau}\right)~, \end{equation} where we have used the abbreviation
$\slash{\cF}_{\mu}\equiv (2\pi/\sqrt{\lambda})\cF_{\mu}$. Since $ \cF^{\mu}dx_{\mu}/d\tau=0$ (which
is merely the statement that no work is done on the quark in its instantaneous rest frame), this
implies that $({d^2\tx/ d\ttau^2})^2=z_m^2 \slash{\cF}^2$, which allows (\ref{tau}) to be simplified
into \begin{equation}\label{tau2} d\ttau=\frac{d\tau}{\sqrt{1-z^4_m \slash{\cF}^2}}~. \end{equation}

Using (\ref{atilde}) and (\ref{tau2}), we can finally rewrite (\ref{recursionsolution2}) in the form
\begin{eqnarray}\label{manyderivatives} \frac{z_m \slash{\cF}^{\mu}}{\sqrt{1-z^4_m
\slash{\cF}^2}}&=&\left(\frac{z^3_m \slash{\cF}^2}{1-z^4_m \slash{\cF}^2}\right){d x^{\mu}\over
d\tau}+\sqrt{1-z^4_m \slash{\cF}^2}{d\over d\tau}\left[\sqrt{1-z^4_m \slash{\cF}^2}{d x^{\mu}\over
d\tau}\right]\\ {}&{}&-z_m \sqrt{1-z^4_m \slash{\cF}^2}{d\over d\tau}\left[\sqrt{1-z^4_m
\slash{\cF}^2}{d\over d\tau}\left[\sqrt{1-z^4_m \slash{\cF}^2}{d x^{\mu}\over
d\tau}\right]\right]+\ldots\nonumber~. \end{eqnarray} This equation of motion for the quark contains
an infinite number of derivatives of $x^{\mu}$, precisely as one would expect for an extended color
charge distribution, based on the classical or quantum electrodynamic analogs
\cite{rohrlich,jackson,monizsharp} mentioned in the Introduction. Notice that to arrive at this
result we have made no assumption about the profile of the charge distribution. The dual string
dynamics automatically incorporates the physics of this profile, which is codified by the slope
$\vec{s}\equiv \p_z \vec{X}(z_m,t)$ \cite{dragtime}. It would be interesting to explore this
connection in more detail through a calculation of $\expec{\tr F^2(x)}$ and similar observables for
an accelerating quark in vacuum \cite{cg,mo,dampingfield}.

A more manageable form of the equation of motion can be obtained by going back to the second equation in
(\ref{recursion}), 
$$ {d^2 x^{\mu}\over
d\ttau^2}={d^2\tx^{\mu}\over d\ttau^2}+z_m {d^3\tx^{\mu}\over d\ttau^3}~. $$ 
Through (\ref{atilde}),
(\ref{tau2}) and (\ref{zm}), this can be reexpressed as \begin{equation}\label{eom} {d\over
d\tau}\left(\frac{m{d x^{\mu}\over d\tau}-{\sqrt{\lambda}\over 2\pi m}
\cF^{\mu}}{\sqrt{1-{\lambda\over 4\pi^2 m^4}\cF^2}}\right)=\frac{\cF^{\mu}-{\sqrt{\lambda}\over 2\pi
m^2} \cF^2 {d x^{\mu}\over d\tau}}{1-{\lambda\over 4\pi^2 m^4}\cF^2}~, \end{equation} which is the
equation we advertised in the Introduction. Notice that it involves only the four-velocity and
four-acceleration of the quark, so, in going from (\ref{manyderivatives}) to (\ref{eom}), we have
traded an infinite number of higher derivatives for a somewhat more complicated $\cF^{\mu}$
dependence. This is to some extent analogous to the possibility of trading (\ref{al}), (\ref{ald}) or
its non-pointlike generalizations for an integro-differential (nonlocal in the force) equation with
derivatives of $x^{\mu}$ only up to second order \cite{rohrlich,jackson}.
 What is different is that our end result, equation (\ref{eom}), does not involve any nonlocality.
 (It is also highly nonlinear, because it includes effects that are neglected for simplicity in
 nearly all previous analyses of radiation damping.)

Before proceeding with the analysis of (\ref{eom}), we would like to note for future use that the
information we have gathered in the process of its derivation, and more specifically, equations
(\ref{recursion}) and (\ref{atilde}), allow Mikhailov's solution (\ref{mikhsol}) to be rewritten purely in
terms of $z=z_m$ data as \begin{equation}\label{mikhsolzm} X^{\mu}(\tau,z)=\left(z-z_m
\over{\sqrt{1-z^4_m \slash{\cF}^2}}\right)\left({dx^{\mu}\over
d\tau}-z^2_m\slash{\cF}^{\mu}\right)+x^{\mu}(\tau)~. \end{equation}

A first check on (\ref{eom}) is to note that it correctly reduces to $m d^2
x^{\mu}/d\tau^2=\cF^{\mu}$ in the pointlike limit $m\to\infty$ (where the Compton wavelength $z_m\to
0)$.  We will now perform a more substantial check by showing that it also makes firm contact with
the results of \cite{dragtime} at finite $m$.

In \cite{dragtime}, two of us generalized the analysis of Mikhailov \cite{mikhailov} to the case of a
quark with finite mass, concentrating for simplicity on motion along one dimension. We showed that
under such circumstances the total string ($=$ field $+$ quark) energy $E$ and momentum $P$ are no
longer given by (\ref{emikh}) and (\ref{pmikh}), but become \begin{eqnarray}\label{dragtimeEP}
E(t)&=&{\sqrt{\lambda}\over 2\pi}\int_{-\infty}^t \!dt\, {F^2\over
m^2}\left(\frac{1-{\sqrt{\lambda}\over 2\pi m^2}F v}{1-{\lambda\over 4\pi^2
m^4}F^2}\right)+\frac{1-{\sqrt{\lambda}\over 2\pi m^2}F v}{\sqrt{1-{\lambda\over 4\pi^2
m^4}F^2}}\gamma m ~,\\ P(t)&=&{\sqrt{\lambda}\over 2\pi}\int_{-\infty}^t \!dt\, {F^2\over
m^2}\left(\frac{v-{\sqrt{\lambda}\over 2\pi m^2}F }{1-{\lambda\over 4\pi^2
m^4}F^2}\right)+\frac{v-{\sqrt{\lambda}\over 2\pi m^2}F }{\sqrt{1-{\lambda\over 4\pi^2
m^4}F^2}}\gamma m \nonumber ~,\end{eqnarray} with $F$ the external force. These expressions show
that
 for $m<\infty$ the rate (seen inside the integrals) at which energy/momentum is radiated
 by the quark  differs from the Lienard result, and the formulas for the intrinsic energy $E_q$ and
 momentum $p_q$ of the quark (given by the terms that follow the integrals) are similarly
 non-standard. Now, the total momentum $P$  of the string  ($=$ field $+$ quark) changes only due to
 the force that we exert on the endpoint ($=$ quark), so $dP/dt=F$, or, using (\ref{dragtimeEP}),
 \begin{equation}\label{dragtimeeom}
 {\sqrt{\lambda}\over 2\pi} {F^2\over m^2}\left(\frac{v-{\sqrt{\lambda}\over 2\pi m^2}F
 }{1-{\lambda\over 4\pi^2 m^4}F^2}\right)
 +{d\over dt}\left(\frac{m\gamma v-{\sqrt{\lambda}\over 2\pi m}\gamma F }{\sqrt{1-{\lambda\over
 4\pi^2 m^4}F^2}}\right)=F~.
 \end{equation}

 Let us now compare this against our equation of motion (\ref{eom}). For linear motion along
 direction $x^1$, we have $dx^{\mu}/d\tau=\gamma(1,v)$ and $\cF^{\mu}=\gamma(Fv,F)$, so the $\mu=1$
 component of (\ref{eom}) reads
  \begin{equation}\label{eom1d}
 {d\over dt}\left(\frac{m\gamma v-{\sqrt{\lambda}\over 2\pi m}\gamma F }{\sqrt{1-{\lambda\over 4\pi^2
 m^4}F^2}}\right)=\frac{F-{\sqrt{\lambda}\over 2\pi m^2}F^2 v }{1-{\lambda\over 4\pi^2 m^4}F^2}~.
 \end{equation}
 This is in precise agreement with (\ref{dragtimeeom}). Similarly, it is easy to see that the $\mu=0$
 component of (\ref{eom}) yields the expected result $dE/dt=Fv$ with $E$ as in (\ref{dragtimeEP}).

 We have thus verified that our quark equation of motion reproduces the energy/momentum split between
 quark and radiation field previously deduced in \cite{dragtime} for the case of one-dimensional
 motion. {}It becomes clear then that (\ref{eom}) encodes the covariant generalization of this split
 to the case of arbitrary motion--- a result that would have been extraordinarily difficult to obtain
 using the non-covariant approach of \cite{dragtime}. To make this generalization explicit, we
 rewrite our equation in the form
 \begin{equation}\label{eomsplit}
 {d P^{\mu}\over d\tau}\equiv {d p_q^{\mu}\over d\tau}+{d P^{\mu}_{\mbox{\scriptsize rad}}\over
 d\tau}=\cF^{\mu},
 \end{equation}
 recognizing
 \begin{equation}\label{pq}
 p_q^{\mu}=\frac{m{d x^{\mu}\over d\tau}-{\sqrt{\lambda}\over 2\pi m}
 \cF^{\mu}}{\sqrt{1-{\lambda\over 4\pi^2 m^4}\cF^2}}
 \end{equation}
 as the intrisic four-momentum of the quark, and
\begin{equation}\label{radiationrate} {d P^{\mu}_{\mbox{\scriptsize rad}}\over
d\tau}={\sqrt{\lambda}\, \cF^2 \over 2\pi m^2}\left(\frac{{d x^{\mu}\over d\tau}-{\sqrt{\lambda}\over
2\pi m^2} \cF^{\mu} }{1-{\lambda\over 4\pi^2 m^4}\cF^2}\right) \end{equation} as the rate at which
four-momentum is carried away from the quark by chromo-electromagnetic radiation.

Using again the fact that $\cF\cdot \p_{\tau}x=0$, we can immediately deduce from (\ref{pq}) the
mass-shell condition $p_q^2=-m^2$, which shows in particular that $p^{\mu}_q$ is indeed a
four-vector, and so the split $P^{\mu}=p_q^{\mu}+P^{\mu}_{\mbox{\scriptsize rad}}$ defined in
(\ref{eomsplit})-(\ref{radiationrate}) is correctly Lorentz covariant. As we indicated above,
$P^{\mu}_{\mbox{\scriptsize rad}}$
  represents the portion of the total four-momentum stored at any given time in the \emph{purely
  radiative} part of the gluonic field set up by the quark. The remainder, $p^{\mu}_q$, includes the
  contribution of the \emph{near} field sourced by our particle, or in quantum mechanical language,
  of the gluonic cloud surrounding the quark, which gives rise to the deformed dispersion relation
  seen in (\ref{pq}). In other words, $p^{\mu}_q$ is the four-momentum of the `dressed' or
  `composite' quark. All of this is completely analogous to the classical electromagnetic case that
  we briefly reviewed in the Introduction, and in particular, to the covariant splitting of the
  Maxwell tensor achieved in \cite{teitelboim}. It is truly remarkable that the AdS/CFT
  correspondence grants us such direct access to this piece of strongly-coupled non-Abelian physics.

  Now that we have performed some checks on (\ref{eom}) and understood its proper physical
  interpretation, we should consider its implications. As noticed already in \cite{dragtime} for the
  case of linear motion,
  a salient feature of the equation of motion (\ref{eom}), as well as the dispersion relation
  (\ref{pq}) and radiation rate (\ref{radiationrate}), is the presence of a divergence when
  $\cF^2=\cF^2_{\mbox{\scriptsize crit}}$, where
  \begin{equation}\label{Fcrit}
  \cF^2_{\mbox{\scriptsize crit}}={ 4\pi^2 m^4\over\lambda}
  \end{equation}
  is the critical value at which the force becomes strong enough to nucleate quark-antiquark pairs
  (or, in dual language, to create open strings) \cite{ctqhat}.

Let us now examine the behavior of a quark that is sufficiently heavy, or is forced sufficiently
softly, that the condition $\sqrt{\lambda |\cF^2|}/2\pi m^2\ll 1$ (i.e., $|\cF^2|\ll
|\cF^2_{\mbox{\scriptsize crit}}|$) holds. It is then natural to expand the equation of motion in a
power series in this small parameter. To zeroth order in this expansion, we have the pointlike result
$m d^2 x^{\mu}/d\tau^2 =\cF$, as we had already mentioned above. If we instead keep terms up to first
order, we find
  \begin{equation} \label{ouraldreduced}
  m {d\over d\tau}\left( {d x^{\mu}\over d\tau}-{\sqrt{\lambda}\over 2\pi
  m^2}\cF^{\mu}\right)=\cF^{\mu}-{\sqrt{\lambda}\over 2\pi m^2}\cF^2 {d x^{\mu}\over d\tau}~.
  \end{equation}
  In the $\cO(\sqrt{\lambda})$ terms it is consistent, to this order, to replace $\cF^{\mu}$ with its
  zeroth order value, thereby obtaining
  \begin{equation}\label{ourald}
  m \left( {d^2 x^{\mu}\over d\tau^2}-{\sqrt{\lambda}\over 2\pi m}{d^3 x^{\mu}\over
  d\tau^3}\right)=\cF^{\mu}-{\sqrt{\lambda}\over 2\pi}{d^2 x^{\nu}\over d\tau^2}{d^2 x_{\nu}\over
  d\tau^2} {d x^{\mu}\over d\tau}~.
  \end{equation}

  Interestingly, (\ref{ourald}) coincides \emph{exactly} with the Lorentz-Dirac equation (\ref{ald})!
  As expected from the discussion in the preceding paragraphs, on the left-hand side we find the
  Schott term (associated with the near field of the quark) arising from the modified dispersion
  relation (\ref{pq}). On the right-hand side we see the radiation reaction force given by the
  covariant Lienard formula, as expected from the  result (\ref{emikh}) \cite{mikhailov}, which is
  the pointlike limit of the radiation rate (\ref{radiationrate}).
  Moreover, by comparing (\ref{ald}) and (\ref{ourald}) we learn that it is  $z_m=\sqrt{\lambda}/2\pi
  m$ that plays the role of characteristic time/size $t_e$ for the composite quark. As we discussed
  around (\ref{Fsquared}), $z_m$ is the Compton wavelength (i.e., size of the gluonic cloud) of the
  quark,
   which is indeed the natural quantum scale of the problem.

   If we continue with the expansion of (\ref{eom}), the result at second order can be written as
   $$
   m\dring{x}^{\mu}-{\sqrt{\lambda}\over 2\pi}\left(\tring{x}^{\mu}-\dring{x}^{\nu}\dring{x}_{\nu}
   \sring{x}^{\mu}\right)+{\lambda\over 4\pi^2
   m}\left(\qring{x}^{\mu}-3\dring{x}^{\nu}\,\tring{x}_{\nu}\sring{x}^{\mu}-{3\over
   2}\dring{x}^{\nu}\dring{x}_{\nu}\dring{x}^{\mu}\right)=\cF^{\mu}~.
   $$
   We can of course continue this expansion procedure to arbitrarily high order in $\sqrt{\lambda
   |\cF^2|}/2\pi m^2$, and at order $n$ in this parameter, we would obtain an equation with
   derivatives up to order $n+2$. Now, it is interesting to note that, in the case of
   non-relativistic QED, there are actually two different scales that appear in the corresponding
   equation of motion \cite{monizsharp}: while the terms with derivatives of fourth and higher order
   are all characterized by the \emph{quantum} (Compton) radius $\lambda_C=\hbar/m$,
   the third-derivative (i.e., Abraham-Lorentz) term involves only the \emph{classical} electron
   radius $c t_e=e^2/mc^2\ll \lambda_C$. In our strongly-coupled non-Abelian setting, the analogs of
   these two scales happen to coincide. On the one hand, $z_m$
    is analogous to $\lambda_C$ in that it gives the size of the cloud of virtual particles
    surrounding the quark, which as we explained at the end of Section \ref{strings2quarkssec}, is
    set by the meson (and not the quark) mass, $z_m=1/m_{\mbox{\scriptsize mes}}$. On the other hand,
    $z_m$ is also analogous to
    $c t_e$ in that it gives the radius of a charge distribution whose chromoelectrostatic energy
    equals the quark mass $m$, if we take into account the \emph{strong-coupling} form of the
    potential $V(L)\propto \sqrt{\lambda}/L$ \cite{maldawilson}.

  As promised in the Introduction, our full equation (\ref{eom}) is thus recognized as an extension
  of the Lorentz-Dirac equation that automatically incorporates the size $z_m$ of our non-classical,
  non-pointlike and non-Abelian source. The passage from (\ref{ourald}) to (\ref{eom}), which can be
  viewed intuitively as the addition of an infinite number of higher derivative terms $(z_m
  d/d\tau)^n$, has a profound impact on the space of solutions. Here we will limit ourselves to two
  general observations, leaving the search for specific examples of solutions to the next section.
  The first is to notice that, unlike its classical electrodynamic counterparts (\ref{al}) and
  (\ref{ald}), our composite quark equation of motion has no pre-accelerating or self-accelerating
  solutions. That is to say, the behavior of the quark at any given time $\tau$ does not depend on
  $\cF^{\mu}(\tau')$ at $\tau'>\tau$, and, in the (continuous) absence of an external force,
  (\ref{eom}) uniquely predicts
  that the four-acceleration of the quark must vanish. Our second observation, however, is that the
  converse to this last statement is not true: constant four-velocity does not uniquely imply a
  vanishing force.
  We will expand on this in the examples of the next section.

  Notice that, in the  systematic approach from our result (\ref{eom}) to the Lorentz-Dirac equation,
  the pathologies that afflict the latter appear only in the very last step, when the
  $\tring{x^{\mu}}$ term is introduced upon approximating (\ref{ouraldreduced}) by (\ref{ourald}).
  Indeed, it has often been advocated to eliminate these pathologies by a `reduction of order'
  procedure (see, e.g., \cite{ll,ghw}), which treats the damping force as a perturbative correction
  and thereby justifies the replacement of (\ref{ourald}) by (\ref{ouraldreduced}). (A closely
  related line of reasoning can be found in \cite{rohrlich2}.) It is therefore satisfying to see that
  the dressed quark equation of motion predicted by AdS/CFT, Eq.~(\ref{eom}), automatically comes out
  in `reduced order' form.

It is curious to note that (\ref{eom}), which incorporates the effect of radiation damping on the
quark, has been obtained from (\ref{mikhsol}), which does \emph{not} include such damping for the
string itself. The supergravity fields set up by the string are of order $1/N_c^2$, and therefore
subleading at large $N_c$. Even more curious \cite{cg} is the fact that it is precisely these
suppressed fields that encode the gluonic field profile generated by the quark, as has been explored
in great detail (mostly at finite temperature) in recent years (see, e.g., \cite{gubsertmunureview}
and references therein). It would be interesting to explore how the split into near and radiation
fields is achieved from this perspective, but we leave this problem to future work
\cite{dampingfield}.

  It is natural to inquire whether the equation of motion (\ref{eom}) can be encoded in a variational
  principle. Since the complete system includes the composite quark in interaction with its radiation
  field, and the four-momentum of the latter is given by an integral over the quark worldline, at
  least naively we would expect that, if it is at all possible to write down an action, it should
  depend bilocally on the worldline. This would rule out the obvious candidate action, namely
  $S=S_{\mbox{\scriptsize NG}}+S_{\mbox{\scriptsize F}}$ with the classical solution
  (\ref{mikhsoltilde}) plugged in. Indeed, the latter procedure simply leads again to
  (\ref{relativisticparticle}) (with $\tau$  in the Nambu-Goto term relabeled to $\ttau$), which
  (\ref{tau}) allows to be converted into
  \begin{equation}\label{relativisticnonparticle}
S= -m\int {d\tau\over\sqrt{1-{\lambda\over 4\pi^2 m^4}\cF^2}}~+\int
d\tau\,A_{\mu}(x(\tau))\,\sring{x^{\mu}}(\tau) %\int d\tau\,\cF_{\mu}(\tau)x^{\mu}(\tau)~.\nonumber
\end{equation}
  Unlike what happened in the
   pointlike (and no radiation damping) limit $m\to\infty$, in this case variation of $S$ with
   respect to $x^{\mu}$ holding $\cF^{\mu}$ fixed does \emph{not} yield the correct equation of
   motion (\ref{eom}). Of course, $S$ is by definition the correct on-shell action for the system, in
   the sense that it yields the right number when evaluated on a given worldline (using the
   $\cF^{\mu}(\tau)$ obtained by solving (\ref{eom})), but it would somehow need to be rewritten in
   bilocal form to explicitly show all relevant $x^{\mu}$ dependence and thus constitute the desired
   variational principle.

   \section{Examples} \label{examplessec}
   \subsection{One-dimensional motion} \label{onedimsubsec}
   We have already seen in Section \ref{eomsubsec} that, in the case of one-dimensional motion, our
   general equation of motion (\ref{eom}) reduces to (\ref{eom1d}). The latter can be further
   simplified to
   \begin{equation} \label{alinear}
a=\frac{z_m\slash{F}(1-v^2)^{3/2}}{\sqrt{1-z_m^4{\slash{F}}^2}}
+\frac{z^2_m\dot{\slash{F}}(1-v^2)}{1-z_m^4{\slash{F}}^2}~. \end{equation} where we have preferred to
express the prefactors in terms of the quark Compton wavelength $z_m$ instead of its mass $m$, and as
in (\ref{atilde}) we have used the abbreviation $\slash{F}\equiv (2\pi/\sqrt{\lambda})F$. Here we see
directly that, in contrast with the usual case, the acceleration at any given time depends not only
on the value of the applied force but also on its rate of change.

It follows from (\ref{alinear}) that a quark that is free for any extended period of time will not
accelerate. In other words, as we had already indicated in the general discussion below
(\ref{ourald}), there are no self-accelerating solutions, which is just as one would have expected
given the extended nature of the charge. On the other hand, the very fact that the charge has a
`deformable' internal structure implies that there is more than one way to get it to follow any given
trajectory. Indeed, for any choice of $v(t)$, (\ref{alinear}) fixes the applied force $\slash{F}(t)$
not algebraically, but through a differential equation that inevitably gives rise to a one-parameter
family of solutions (differing by their initial conditions).

The simplest example of this non-uniqueness of the external force is the case of constant velocity,
where (\ref{alinear}) can be easily seen to imply that \begin{equation}\label{constantvelocity}
\slash{F}(t)=\pm{1\over z_m^2}\mathrm{sech}\!\left({\sqrt{1-v^2}\over z_m}(t-t_0)\right)~,
\end{equation} with $t_0$ an integration constant. Only for $t_0\to\pm\infty$ does one recover the
simple result $F(t)=0$. For all finite values of the integration constant, the force starts out at
zero at asymptotically early times, rises steadily until it attains the critical value
$F=\sqrt{\lambda}/2\pi z_m^2$ given by (\ref{Fcrit}) at $t=t_0$, and then approaches zero again as
$t\to\infty$.

It is certainly peculiar that one can continually apply a force to the quark and still have it move
at constant velocity. Nonetheless, it is easy to verify through numerical integration that indeed the
application of the force (\ref{constantvelocity}) to the endpoint of a string whose initial profile
is chosen in accord with (\ref{mikhsolzm}) produces no acceleration. The energy provided to the
system by $F(t)$ does not translate into an increase of the endpoint velocity, but into a continuous
modification of the string tail, or, in gauge theory language, a change of the gluonic field profile.
In fact, with the formulas derived in the previous section, we can make a much more precise
statement: (\ref{pq}) and (\ref{radiationrate}) reduce to \cite{dragtime} $$
E_q=\left(1-z^2_m\slash{F}v \over \sqrt{1-z_m^4 \slash{F}^2}\right)\gamma m~,\qquad
{dE_{\mbox{\scriptsize rad}}\over dt}=z_m^2 \slash{F}^2\left(1-z^2_m\slash{F}v \over 1-z_m^4
\slash{F}^2\right)~, $$ where we see that application of the force (\ref{constantvelocity}) results
in \begin{eqnarray} E_q&=&\left\{\left|\mathrm{coth}\!\left({t-t_0\over \gamma z_m}\right)\right|\mp
v\left|\mathrm{csch}\!\left({t-t_0\over \gamma z_m}\right)\right|\right\}\gamma m~,\nonumber\\
{dE_{\mbox{\scriptsize rad}}\over dt}&=&\mathrm{csch}^2\!\left({t-t_0\over \gamma
z_m}\right)\left[1-v\,\mathrm{sech}^2\!\left({t-t_0\over \gamma z_m}\right)\right]~,\nonumber
\end{eqnarray} which show precisely what fraction of the energy goes into (or comes out from)
rearranging the near gluonic field, and what fraction goes into radiation. In the limit
$t_0\to\pm\infty$, no force is applied and we of course recover $E_q=\gamma m$,
$dE_{\mbox{\scriptsize rad}}/ dt=0$ at all times. Similar conclusions can be drawn about the linear
quark momentum.

Notice that the width of the time interval over which the force (\ref{constantvelocity}) differs
appreciably from zero is just the Compton wavelength of the quark, $z_m$ (with an appropriate Lorentz
dilation factor), which serves to emphasize that this peculiar phenomenon is made possible only due
to the non-pointlike nature of our non-Abelian source.

This non-uniqueness of the force is equally present for arbitrary trajectories.
 Another example that is easy to study is the case of constant force. If we assume that
 $F(t)=\,$constant, (\ref{alinear}) reduces to $d(\gamma
 v)/dt=z_m\slash{F}/\sqrt{1-z_m^4\slash{F}^2}$, which has the same form as the standard equation of
 motion for a particle with constant proper acceleration, except that the force here is non-linearly
 rescaled. The solution is thus
$$ x(t)=x_0\pm\sqrt{{1\over z_m^2\slash{F}^2}-z_m^2+(t-t_0)^2}~, $$ where $x_0$ and $t_0$ are
integration constants. But if we run the argument in reverse, and ask what type of force would lead
to the hyperbolic motion $x(t)=x_0\pm\sqrt{C^2+(t-t_0)^2}$, we see again that the differential
equation (\ref{alinear}) admits solutions other than the obvious
$\slash{F}(t)=1/z_m\sqrt{C^2+z_m^2}$.

Before closing this subsection, we would like to make one additional observation. After some
straightforward algebra, it is easy to see that (\ref{radiationrate}) can be rewritten in the form
\begin{equation}\label{radiationrate2}
 {d P^{\mu}_{\mbox{\scriptsize rad}}\over d\tau}={\sqrt{\lambda}\,\over 2\pi
 m^2}\left[\frac{\cF^2}{\sqrt{1-{\lambda\over 4\pi^2 m^4}\cF^2}}\right]p_q^{\mu}~.
\end{equation} Plugging this into (\ref{eomsplit}), one obtains a first order differential equation
for $p_q^{\mu}$ with the same structure as the Langevin equation, but with a friction coefficient
that depends on the external force, \begin{equation}\label{langevintype}
 {d p_q^{\mu}\over d\tau}=-\mu(\cF)p_q^{\mu}+\cF^{\mu}~,
\end{equation} where $\mu(\cF)\equiv(\sqrt{\lambda}/ 2\pi m^2)\cF^2/\sqrt{1-{\lambda/4\pi^2
m^4}\cF^2}$. In the case of motion in one spatial dimension, this equation can be solved
analytically, \begin{multline}
 p_q=\exp\left({-{\sqrt{\lambda}\,\over 2\pi m^2}\int^t{\frac{\cF^2(x)}{\sqrt{1-{\lambda\over 4\pi^2
 m^4}\cF^2(x)}}dx}}\right)\\ \times \left[A+ \int^t{\exp\Bigg({{\sqrt{\lambda}\,\over 2\pi
 m^2}\int^y{\frac{\cF^2(x)}{\sqrt{1-{\lambda\over 4\pi^2 m^4}\cF^2(x)}}dx\Bigg)\cF(y)dy}}}\right]~,
\end{multline} with $A$ an integration constant. We should stress that this solution is valid only in
the case of one dimensional motion, where ${\cal F}^2=\vec{F}^2$. In two or three spatial dimensions
we have not been able to construct an explicit solution, because in that case ${\cal F}^2$ involves
the velocity of the quark, and remains undetermined until we find the quark trajectory. Nevertheless,
it is interesting to note that, in the case of constant $\cF^2$ (which is \emph{not} the same as
constant $\vec{F}^2$), equation (\ref{langevintype}) can be interpreted as a Langevin equation of
motion for a particle with momentum $p_q$ (up to a constant term). The force in terms of the physical
time $t$ is just the force needed to move the quark in a `dissipative medium' characterized by the
constant friction coefficient $\mu$.

   It would be interesting to explore the dynamics of the dressed quark in two or three spatial
   dimensions. However, as we have just remarked, in that case (\ref{eom}) is highly nonlinear not
   only in the external force (which it was already in one dimension), but also in the velocity of
   the quark. For this reason, it is unfortunately very difficult to find analytic solutions to it.
   In the heavy quark (or small force) approximation where (\ref{eom}) linearizes and reduces to the
   Lorentz-Dirac equation (\ref{ourald}), we could of course carry over to our setting the various
   solutions that have been worked out in the past. For instance,  in the first reference of
   \cite{rohrlich}, Rohrlich was able to find an analytic solution for a central force problem using
   the Frenet equations. Starting from this and perturbing the solution it should be possible to
   obtain (at least numerically) the first correction beyond Lorentz-Dirac predicted by our
   framework, and deduce for instance the r
 ate of synchrotron radiation. Such a calculation might shed some additional light on the physics
 behind these extended objects, but is not central for the purposes of our analysis here, so we
 prefer to leave it for future work.

   \subsection{Nonrelativistic limit} \label{nrsubsec}

   Let us now choose a specific Lorentz frame $(\vec{x},t)$, and restrict attention to motions such
   that the quark velocity $\vec{v}\equiv d\vec{x}/dt$, acceleration $\vec{a}\equiv d^2\vec{x}/dt^2$,
   jerk $\vec{\jnodot}\equiv d^3\vec{x}/dt^3$, and all higher derivatives, as well as the force
   $\slash{F}$ and its rate of change $d\slash{F}/dt$, are small in units of the Compton wavelength
   $z_m$. Under such conditions $d\tau\simeq dt$, and the spatial components of
   (\ref{manyderivatives}) adopt the linearized form
   \begin{equation}\label{manyderivativesnr}
   {d^2\vec{x}\over dt^2}-z_m {d^3\vec{x}\over dt^3}+z_m^2 {d^4\vec{x}\over
   dt^4}-\ldots=z_m\vec{\slash{F}}~.
   \end{equation}
   Adding to this expression its time derivative multiplied by $z_m$, we obtain the simplified form
   \begin{equation}\label{eomnr}
   {d^2\vec{x}\over dt^2}=z_m\vec{\slash{F}}+z^2_m{d\vec{\slash{F}}\over dt}~,
   \end{equation}
   which is the linearized version of (\ref{eom}). We see here very directly that, as explained in
   Section \ref{eomsubsec}, the unusual dependence on the rate of change of the force encodes an
   infinite number of higher derivatives of the quark trajectory $\vec{x}(t)$, derivatives which in
   turn reflect the extended nature of the quark.

   It is also useful to note that in this linearized limit, the string embedding (\ref{mikhsolzm})
   can be rewritten in the form
   \begin{equation}\label{mikhsolnr}
   \vec{X}(t,z)=\vec{x}(\tret)+(z-z_m)\sum_{l=1}^{\infty}(-z_m)^{l-1}
   {d^l\vec{x}\over dt^l}(\tret)~,\quad \tret\equiv t-z+z_m~,
   \end{equation}
   which involves only the quark worldline $\vec{x}(t)$ and not the force $\vec{F}(t)$. If we plug
   this into the (quadratic version of the) total string energy $E(t)\equiv-\int dz\Pi^t_t=\int dz
   (\dot{\vec{X}}^2+\vec{X}^{'2})/2$, and imitate Mikhailov's procedure \cite{mikhailov} (see Section
   \ref{eomsubsec}), the expression naturally splits into
   $$
   E(t)=\int_{-\infty}^t d\tret {dE_{\mbox{\scriptsize rad}}\over d\tret}+E_q(t)~,
   $$
   with
   $$
   E_q(t)={1\over 2}m\left(\sum_{l=1}^{\infty}(-z_m)^{l-1} {d^l\vec{x}\over dt^l}(t)\right)^2
   $$
   and
   $$
   {dE_{\mbox{\scriptsize rad}}\over d\tret}=\left(\sum_{l=2}^{\infty}(-z_m)^{l-2} {d^l\vec{x}\over
   dt^l}(\tret)\right)^2+\frac{z_m}{(t-\tret+1)^2}{d\vec{x}\over dt}(\tret)\cdot
   \left(\sum_{l=2}^{\infty}(-z_m)^{l-2} {d^l\vec{x}\over dt^l}(\tret)\right)~.$$
   Using (\ref{manyderivativesnr}), it is easy to check that these expressions indeed coincide with
   the non-relativistic limit of the time component of (\ref{pq}) and (\ref{radiationrate}). The
   spatial components can be verified in a similar manner. Notice that, in the relativistic case
   analyzed in \cite{dragtime} and Section \ref{eomsubsec} of the present work, the explicit
   presence of the force $\mathcal{F}^{\mu}$ in the solution (\ref{mikhsolzm}) prevents
   us from deriving the split (\ref{eomsplit}) as we did here, directly imitating Mikhailov's
   procedure.
   It is therefore comforting to see that, at least in the non-relativistic case, the direct result
   at $z=z_m$ indeed agrees with what we had previously inferred indirectly from Mikhailov's
   unambiguous split for the auxiliary data at $z=0$.

   Just like in the fully relativistic setting, if we directly substitute (\ref{mikhsolnr}) into the
   (quadratic version of the) string action (\ref{nambugoto}) $+$ (\ref{externalforce}), we \emph{do
   not} arrive at a variational principle that correctly encodes the equation of motion
   (\ref{manyderivativesnr}) or (\ref{eomnr}). Here, however, it is easy to write down an action that
   does the right job. In fact, we have not one but (at least) two options: we can evidently get
   (\ref{manyderivativesnr})  from
   $$
   S_{\mbox{\scriptsize nr}}=-{m\over 2}\int dt\left[\left({d\vec{x}\over dt}\right)^2
   +\left({d^2\vec{x}\over dt^2}\right)^2+\left({d^3\vec{x}\over dt^3}\right)^2
   +\ldots\right]+\int dt \vec{F}\cdot\vec{x}~,
   $$
   and (\ref{eomnr}) evidently follows from
   $$
   S'_{\mbox{\scriptsize nr}}=-{m\over 2}\int dt\left({d\vec{x}\over dt}-z_m\int^t
   dt'\vec{\slash{F}}(t')-z_m^2\vec{\slash{F}}(t)\right)^2~.
   $$

   It is curious to note that the action $ S_{\mbox{\scriptsize nr}}$ in terms of higher order
   derivatives of the position vector with
respect to $\tau$ is reminiscent of the dynamical description of a particle in noncommutative
symplectic mechanics given in \cite{nachoto}, where it was found that the action can be written in
terms of higher order derivatives of the position vector despite the fact that the equations of
motion are of second order.

   Let us now explore some solutions. Either from $S'_{\mbox{\scriptsize nr}}$ or directly from the
   equation of motion (\ref{eomnr}), one immediately has a first integral of the motion,
   $$
   {d\vec{x}\over dt}=z_m\int_{-\infty}^t
   dt'\vec{\slash{F}}(t')+z_m^2\vec{\slash{F}}(t)+\vec{v}_{-\infty}~.
   $$
   The case with zero acceleration corresponds to ${\slash{\vec F}}+z_m\dot{{\slash{\vec F}}}=0$,
   whose solution is $\vec{\slash{F}}(t)=
   \vec{\slash{F}}_0\exp[-(t-t_0)/z_m]$. This is of course simply the non-relativistic limit of
   (\ref{constantvelocity}). Since we are working now in the linearized approximation, this solution
   of the homogeneous equation of motion can be added to any solution of the inhomogeneous equation
   (\ref{eomnr}) to yield another solution, so it is very clear that the non-uniqueness of the force
   is present for any quark trajectory whatsoever. The reason is by now familiar to us: the energy
   provided to the dressed quark by this one-parameter family of forces has the effect of modifying
   the gluonic field profile, and consequently does not translate into an increase of velocity.

\section*{Acknowledgements}

We are grateful to David Mateos, Mat\'\i as Moreno and Miguel \'Angel P\'erez for useful
conversations, and to Juan Felipe Pedraza for spotting an error in an early draft of this paper. This
work was partially supported by Mexico's National Council of Science and Technology (CONACyT) grant
50-155I, as well as by DGAPA-UNAM grant IN116408.

\end{document}